# A 6 kV arbitrary waveform generator for the Tevatron lens[*]

**H Pfeffer, G Saewert[†]**

*Fermi Nation Accelerator Laboratory,
PO Box 500, Batavia, IL, 60510, U.S.A
E-mail*: saewert@fnal.gov

ABSTRACT: This paper reports on a 6 kV modulator built and installed at Fermilab to drive the electron gun anode for the Tevatron Electron Lens (TEL).  The TEL was built with the intention of shifting the individual (anti)proton bunch tunes to even out the tune spread among all 36 bunches with the desire of improving Tevatron integrated luminosity.  This modulator is essentially a 6 kV arbitrary waveform generator that enables the TEL to define the electron beam intensity on a bunch-by-bunch basis.  A voltage waveform is constructed having a 7 µs duration that corresponds to the tune shift requirements of a 12-bunch (anti)proton beam pulse train.  This waveform is played out for any one or all three bunch trains in the Tevatron.  The programmed waveform voltages transition to different levels at time intervals corresponding to the 395 ns bunch spacing.  Thus, complex voltage waveforms can be played out at a sustained rate of 143 kHz over the full 6 kV output range. This paper describes the novel design of the inductive adder topology employing five transformers.  It describes the design aspects that minimize switching losses for this multi-kilovolt, high repetition rate and high duty factor application.

---

[*] Work supported by Fermi Research Alliance, LLC under Contract No. DE-AC02-07CH11359 with the US Department of Energy.
[†] Corresponding author.

# Contents

## 1. Background and introduction

### 1.1 Background

Beam-beam effects related to Tevatron luminosity are described in [1]. One of the measures anticipated to counteract adverse beam-beam effects was to implement bunch-by-bunch tune shifting of individual (anti)proton bunches by use of the Electron Lenses installed in the Tevatron [2]. To be to most effective, the TEL would need to be able to shift the tunes of all 36 bunches to even out the tune spread. This modulator was designed to drive the electron gun anode for this task and was installed in the fall of 2008 into TEL2.

### 1.2 Modulator requirements

The Tevatron bunch structure determines the modulator's timing requirements. The tune shift range in addition to the gun purveyance determine the necessary modulator output voltage range. By the year 2008, it was determined that even 5 kV would be adequate. Internal power dissipation is certainly a limiting factor when switching to kilovolts at these rep rates and high duty factors. Therefore, understanding the timing issues provides an overview of the rationale for the modulator design.

There are 36 bunches of both protons and antiprotons in the Tevatron traveling in opposite directions. Both particle beams contain three bunch trains containing 12 bunches followed by a 2.6 µs abort gap. Bunches are spaced 395 ns apart. It was determined that the pattern of shifted tunes was nearly the same for all three trains. Therefore, the modulator was designed to store information to produce a single voltage waveform to compensate uneven tunes in one 12-bunch pulse train. This waveform can then be triggered for any one of the three (anti)proton bunch trains, or it can be triggered for all three continuously and thereby even out the tunes of all 36

Tevatron bunches. The tune shift pattern varies from one Tevatron store to another, and the stored waveform in the modulator can be updated remotely over Ethernet at any time.

Adequate resolution for equalizing the tune shifts can be satisfied with a fixed number of discrete voltages and need not be infinitely adjustable. For this reason the modulator's output resolution of "arbitrary" voltages is 16 discrete voltage levels from minimum to maximum. The modulator output steps to different values and must settle to a reasonable level by the time of arrival of a Tevatron beam bunch. The modulator outputs the minimum (or the lowest) voltage for those bunches whose tunes are not to be shifted and is always the minimum value during the abort gap. Figure 5 is an example of an output waveform compensating a 12-bunch train.

This modulator was designed to output complex waveforms over a 6.4 kV range. Absolutely, the output voltage will transition to the maximum—in some zigzag pattern—and back to minimum once in each 7 μs waveform period. Also, this waveform can be triggered for each of the three pulse trains. This means that voltages will transition from minimum to maximum at 143 kHz. However, existing tune shift patterns are more complicated than that. To even out the tune spread of one 12-bunch train requires the capability of going from minimum to maximum and back not once but as many as three times within each 7 μs waveform period. This means the modulator would have to be able to switch fully on and off at a sustained average rate of 429 kHz to even out all 36 Tevatron bunches. This is what is meant by "worst case switching".

## 2. Modulator design

Figure 1 is the block diagram of the modulator showing the major circuit components and system connections. The modulator is referenced to ground, but its output is capacitively coupled to the electron gun anode that is DC biased to the cathode. The heart of the modulator is the voltage adder comprised of transformers T1-5. The transformer primaries are driven by H-bridges independently, but their voltages are summed through the series-connected secondaries.

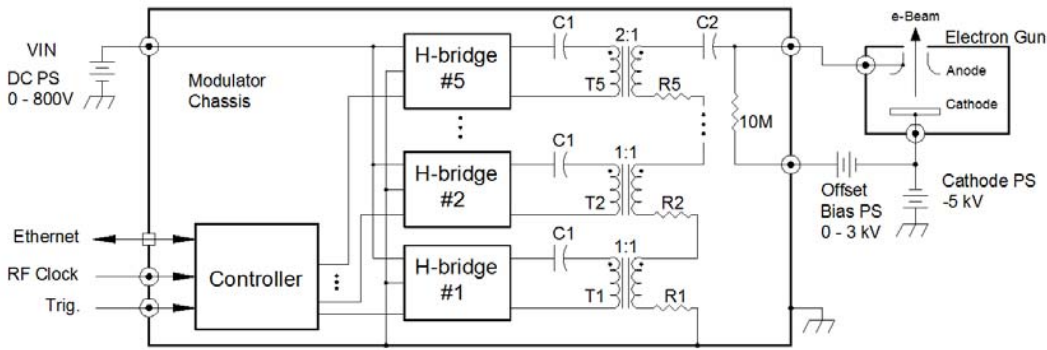

**Figure 1.** Simplified modulator block diagram and external connections.

This modulator was designed to drive the gun anode with positive voltage—anode to cathode. Positive anode voltage $Va$ generates electron current $Ie$ following the expression: $Ie = kVa^{3/2}$, where $k$ is the gun's purveyance. Any negative anode to cathode voltage cuts the gun off. This modulator's output is AC coupled out of necessity because of the use of transformers, but can be DC shifted at the gun anode. Voltage waveforms of arbitrary amplitude are constructed by a combination of two things: (1) setting the value of DC voltage VIN to define the dynamic voltage range and (2) controlling the bipolar drive of the H-bridges to define the waveform pattern. The DC power supply VIN is external to the modulator and is independently controllable over the continuous range of 0 – 800 V DC. Once set, VIN remains fixed.

The bipolar drive capability of H-bridges is utilized. The H-bridges are always switched to one of three states: positive, negative or zero. The positive state occurs when the H-bridge applies +VIN to the primary. When the H-bridge is switched to the negative state –VIN is applied to the primary. The H-bridge shorts out the primary in the zero state. Note that the normally-off, or default, switch state is negative. This enables the modulator to be used conceptually as a unipolar generator that outputs positive voltage waveforms. Each transformer contributes a positive voltage when the H-bridge is switched to either of the two states other than the negative state. In other words, the transformers are driven with either 0, 1×VIN or 2×VIN.

The magnitude of voltage contributed by each transformer depends on its turns ratio. Transformers T1-3 have a turns ratio of 1:1 and T4 and T5 are wound 2:1. The modulator output is simply the sum of secondary voltages, and the maximum waveform peak-to-peak voltage the modulator will output for a given VIN setting is

$$Vo\_\max = 2 \times VIN(3 + 2 \times (1/2)) = 8 \times VIN \quad [Vpp].$$

For example, *Vo_max* is 6.4 kV when VIN is 800 V DC and all H-bridges are switched to "positive". Given that there are two different turns ratios among the five transformers means there are always 16 equally spaced discrete output voltages that can possibly be generated by controlling the H-bridges in various drive combinations. Two different turns ratios were implemented in order to provide better resolution over the output voltage range compared with having all five transformers the same. The number of optional voltages is 16 in this case rather than 10.

In practice, the maximum electron beam current is decided first, from which *Vo_max* is determined. This then determines the needed input voltage VIN. The minimum settable voltage step size within the constructed waveform is *Vo_max*/16 (or VIN/2).

The H-bridges are controlled independently and can be switched from any state to any state. H-bridges that will be changing state do so at the same time and synchronous with the RF clock at the defined 395 ns spaced, beam bunch time slots. The resulting output voltage step-changes to the arbitrary voltage to produce the desired electron current for each (anti)proton beam bunch.

The capacitors C1 shown in Figure 1 play an important role. First of all, they provide DC blocking to allow the H-bridges to output DC voltage VIN indefinitely. Secondly, as switching commences, a voltage develops across C1 that results in the voltage across the transformer primary to automatically average to zero volts—regardless of the H-bridge switching pattern duty factor. The transformers will not march towards saturation. This allows the generation of truly arbitrary waveforms to be generated.

Capacitor C2 enables the output to be DC offset by way of the Offset Bias PS making full use of AC coupled, high duty factor waveforms to generate electron beam. The Offset Bias PS voltage can be made to shift the anode voltage such that the waveform's minimum voltage is equal to the cathode voltage. When this is done, all waveform voltage above the minimum will generate electron beam. Use of the Offset Bias PS is optional. As long as enough voltage can be generated at the anode to produce desired beam current, it doesn't matter how much the anode is driven below the cathode; and low duty factor waveforms do not result in much DC shift anyway.

## 3. Transformer design

Crucial to the modulator performance is the transformer design. Transformer parasitics define the modulator's pulse response characteristics. These effects only compound as multiple transformers are stacked in series. Requirements are that the modulator output settles to a reasonably flat in ~400 ns. Figure 2 shows the secondary side parasitics SPICE model used in development.

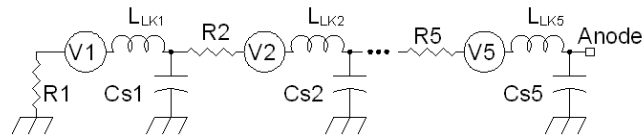

**Figure 2.** Transformer secondary side equivalent circuit model. $L_{LK}$'s are the transformer leakage inductances reflected to the secondary side. The secondary voltages are represented by the V's. Cs's are the secondary stray capacitances and the resistors, labeled as appearing in Figure 1, are added to dampen the pulse response.

Tests were made to verify a good pulse transformer parasitics model. Pulse responses were compared between SPICE simulations and those of driving a wound transformer. Measured transformer parasitics were put into the SPICE model for a single RLC section of the Figure 2 circuit. Comparison was made by adjusting dampening resistor values until both circuits were critically damped. (The load was open circuit.) The result was that SPICE's pulse response rise time was only about 20% slower than the real circuit. Thus, SPICE modeling was effective in guiding the iterative design process.

The expression used for calculating secondary leakage inductance of the transformer design is

$$L_{LK} = \frac{4\pi \times 10^{-7} A N_P^2}{\lambda} \quad \text{(H) in mks units,}$$

where $N_P$ is the number of primary turns, $A$ is area between primary and secondary coils and $\lambda$ is the height of the primary winding [3]. The primary and secondary winding heights are equal, and both are single layer. Layers of .010 inch mica impregnated Nomex® 418 insulation and the secondary winding are wound tightly over the primary winding on all transformers. Also, the secondary turns of T1-3 are located directly over the primary windings of these 1:1 transformers to help reduce leakage inductance. As suggested in the above expression, spreading apart the windings to increase the winding height also lowers leakage inductance. This was found to be true except to a winding separation of about 0.15 inches beyond which little more is gained. The need for increased dielectric strength increases with each transformer, so the number of insulation layers progresses from T1 to T5.

The expression used for calculating stray secondary to primary capacitance with a turns ratio 1:1 is

$$Cs = \frac{\varepsilon S \lambda}{d} \quad (F),$$

where $\varepsilon$ is the dielectric constant of the medium between primary and secondary, $S$ is the mean circumference of the primary and secondary and $d$ is the distance between primary and secondary [4].

Placing a resistor in series with each transformer secondary, as shown as R1-5 in Figures 1 and 2, very effectively dampens the output pulse response resulting from the parasitics of these five transformers. To achieve critical dampening of the overall circuit, each resistor simply needs to be the value obtained from the expression $R = 2 \times \sqrt{(L_{LK} / Cs)}$ of each transformer. The calculated resistor values were close to each other, however, so a single value was chosen for all five. Once constructed, the resistors needed to be decreased by adding parallel resistors to account for unanticipated parasitic capacitance of the final assembly.

There is a second important role resistors R1-5 play. They dissipate the power from charging to high voltage and discharging all the parasitic capacitance in the transformer secondary circuit. Capacitive currents flow through the resistance in series—composed of these resistors and the Rds-on FET resistance—where $I^2R$ power is dissipated. This average power is calculated from $CV^2f$, where $C$ is each capacitance, $V$ is the capacitor voltage and $f$ is the switching rate. Something less than 5% is dissipated in the H-bridge FETs, so these physical resistors not only dampen the output response, but they dissipate significant switching losses rather than the H-bridge FET's. The parasitics on the output lead is about 140 pF; so for example, a single pulse to maximum voltage once every Tevatron revolution results in 270 W of dissipation in this part of the circuit, while worst case switching could be over 2 kW. (The modulator has never been run at worst case.)

**Table 1.** Measured transformer parameters. Np and Ns are primary and secondary turns. Lp and $L_{LK}$ are primary excitation and leakage inductance. Cs is secondary stray capacitance. $\lambda$ is winding height. d is primary to secondary distance. Vc is secondary winding corona extinction voltage (primary grounded).

| X | Np | Ns | Lp (µH) | $L_{Lk}$ (µH) | Cs (pF) | λ (in.) | d (in.) | Vc (volts) |
|---|----|----|---------|---------------|---------|---------|---------|------------|
| 1 | 11 | 11 | 2130 | 2.13 | 42 | 2.25 | .050 | 1400 |
| 2 | 11 | 11 | 2430 | 2.09 | 42 | 2.25 | .050 | 1600 |
| 3 | 11 | 11 | 2076 | 2.28 | 34 | 2.0 | .060 | 2500 |
| 4 | 10 | 5 | 1520 | 3.79 | 25 | 1.6 | .080 | 3150 |
| 5 | 10 | 5 | 1520 | 3.75 | 22 | 1.6 | .080 | 3000 |

The measured transformer parameters are listed in Table 1 of the final assembly. Estimates of average core flux density were made to better understand core losses and to choose the core size. The design intended to allow for sustained switching at up to 429 kHz. Manganese-zinc ferrite material MN8CX from Ceramic Magnetics, Inc. was chosen. The vendor's expression for core loss of this material is $P' = 5.9 \times 10^{-16} B^{2.5} F^{1.9}$, where $P'$ is power per unit volume in mW/cm$^3$, $B$ is flux density in Gauss and $F$ is switching frequency in Hertz. It was observed by making measurements of a wound sample core that losses were nearly half of what would be expected from this loss expression assuming the magnitude of $B$ is the sinusoid peak when switching at rate $F$.

Given the core loss expression above, core loss $P$ (Watts) relates to the cross sectional area $A$ as: $P \propto 1/A^{3/2}$ for a rectangular core in which the core size does not change much in the direction of the flux path for variations in core area. This shows that core losses decrease at a faster rate than incremental increases in its area. Thus, increasing core area buys not only lower core temperature but fewer primary turns that, in turn, decreases leakage inductance. The core size chosen for all transformers was 2.0 square inches. Actual power dissipation of these transformers was measured for three switching rates and is shown in Figure 4.

## 4. H-bridge design

An H-bridge and transformer are assembled as a module as shown in Figure 3. This minimized lead length and related leakage inductance. The FET heat sinks and transformer core are close together forming a channel through which the forced air is directed for cooling.

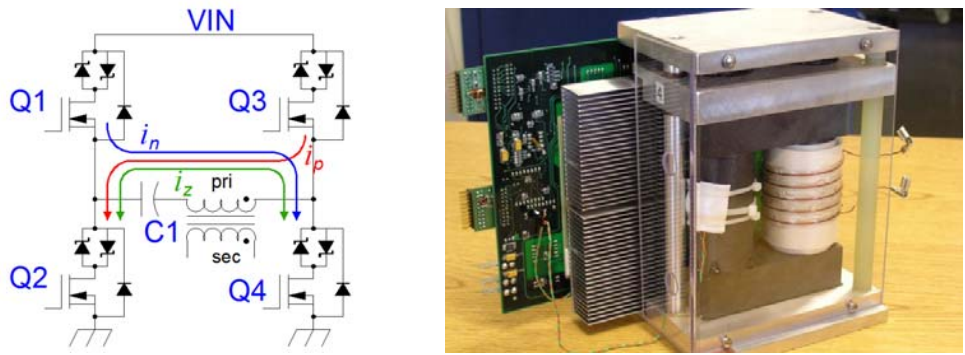

**Figure 3.** H-bridge and transformer assembled module. Diagram shows direction of currents in the positive (ip), negative (in) and zero (iz) switch states.

The FET conducted currents are small, so both $I^2R$ conduction losses as well as transition switching losses are on the order of a few watts each per FET. However, switching losses due to charging and discharging drain to source capacitance are much larger. The FET Rds-on is the only resistance in series with this capacitance to dissipate the stored energy. Each FET dissipates on average $P' = C \times VIN^2 f$ (Watts), where $C$ is the total capacitance across both transistors of a half-bridge and $f$ is the H-bridge switching rate. These switching losses were measured with no load connected to the H-bridge and are shown in Figure 4 on the left.

The MOSFETs are IXZR08N120 from IXYS Corp. chosen for having low $C_{OSS}$ and fast switching characteristics. The dead-time delay was fixed at 10 ns between each FET turn off and the turn on of the other FET in each half-bridge. Extra diodes were added to prevent damaging the FET's when commutating reverse current. These diodes are shown in Figure 3.

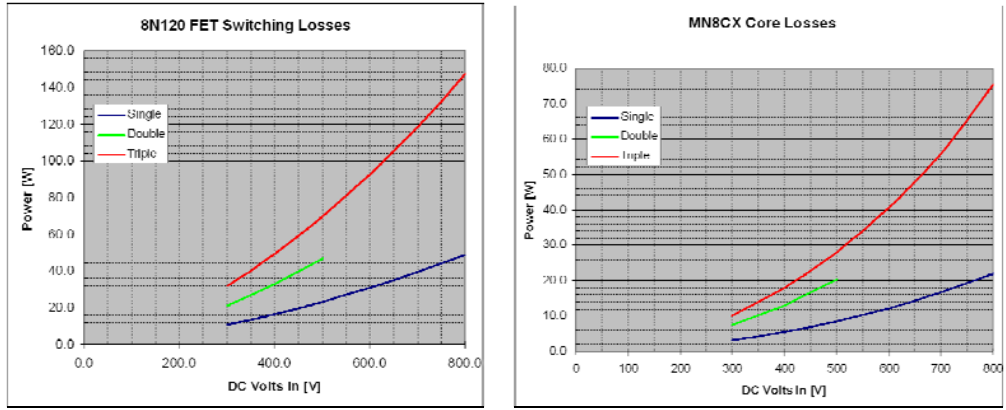

**Figure 4.** Switching losses measured in the H-bridge and the transformers. H-bridge switching losses are evenly distributed among all four FET's. Switching rate "single" is a complete positive to negative switching cycle one time at 143 kHz, "double" is twice at 286 kHz and "triple" is three times at 429 kHz. (The data set is not complete for "double".)

## 5. Performance

Figure 5, left, shows the modulator output of an 790 ns wide pulse produced with VIN set equal to 500 V. All H-bridges transition together from negative to positive and back to negative to produce it. The 10-90% rise time is 200 ns.

A much more complex waveform is shown on the right that was programmed to compensate the 12-bunch trains of Tevatron store #5162. Voltage for each bunch was chosen that generates a desired electron beam current to shift that bunch's tune. The waveform is repeated with bunch 13, etc., showing the single waveform used to compensate back-to-back Tevatron 12-bunch trains. VIN was set to 600 V, and the maximum voltage produced, anode to cathode, is 4800 V at bunch 12. Anode to cathode voltage of zero occurs at bunch 5 as well as during the abort gap between bunches 12 and 13. This waveform is shown without the bias offset voltage applied at the anode.

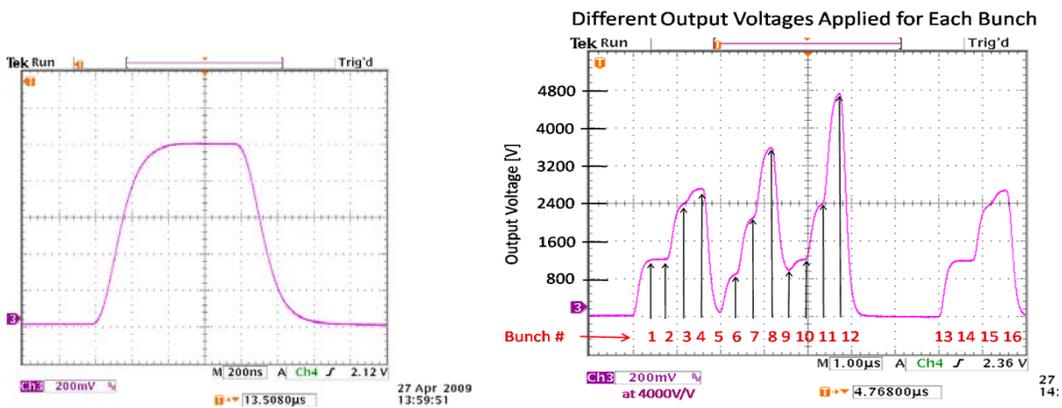

**Figure 5.** Modulator output response plots. Left is a single pulse, 790 ns wide with VIN equal to 500 Vdc. Right is the response to compensate each Tevatron bunch. A 2.6 µs abort gap is between bunch 12 and 13. The arrows show the alignment of the antiproton bunches with the voltage waveform.

## 6. Construction

A chassis was specially designed for the modulator. H-bridge/transformer modules are enclosed in shielded compartments that allow forced air to flow in a directed manner across the FET's, transformer cores and secondary side resistors. One 300 cfm fan forces air throught the chassis. The inside of the chassis behind the front door is shown on the left of Figure 6. The five H-bridge/transformer modules are behind the panels with holes. These panels shield the low level trigger signals orginating from the controller out of view on the left of the chassis.

It is worth mentioning that gate drive isolation was accomplished using ADuM1100 *i*Coupler digital isolators from Analog Technology and/or Si8844x digital isolators from Silicon Laboratories. These fast isolators allow for low jitter in preserving nano-second timing across 800 V and have high common mode transient immunity greater than 25 kV/μs.

On the right of Figure 6 are the secondary side components in the rear of the chassis—between another shielding panel and the rear door. This separate shielded region houses resistors R1-5 and capacitor C2 as labeled in Figure 1. Observable is the manner that these resistor values were reduced in value to tune the circuit for critically dampening by the addition of resistor strings across the 200 Watt Globar resistors. Also labeled are transformers T1-5 located on the other side of the panel.

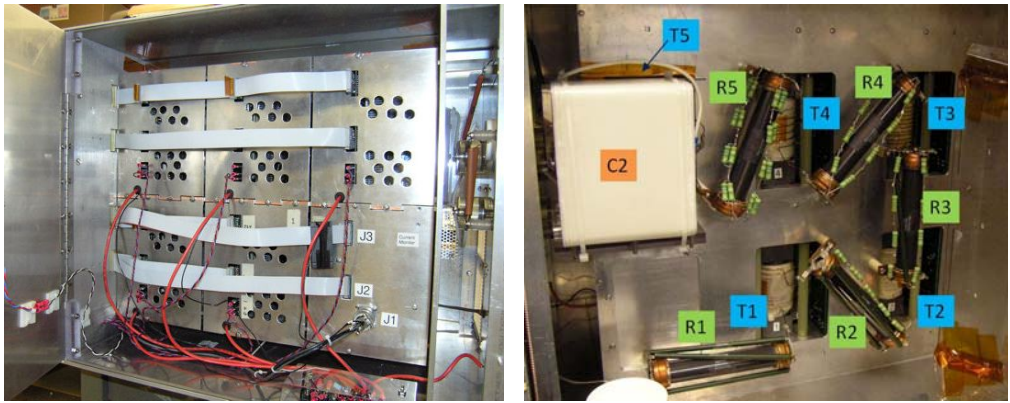

**Figure 6.** Photos of the inside front, on the left, and rear chassis, on the right.

## 7. Conclusion

This modulator demonstrates the effectiveness of driving transformers in a bipolar manner in an inductive adder topology to generate high rep rate, high duty factor, kilo-volt waveforms. Capacitors in series with the transformer primaries prevent transformers from saturating and allows truly arbitrary waveforms to be generated. Employing transformers having different turns ratios increases resolution of selectabe output values.

## 8. Acknowledgments

The authors wish to thank Jeff Simmons for his significant contribution of electrical technician work, Kevin Roon for working out the mechanical design details, Dan Wolff for support of departmental resources and Vladimir Shiltsev for being the enthusiastic, motivated leader of the TEL project.

## References

[1] V. Shiltsev et al., *Beam-beam effects in the Tevatron*, Phys. Rev. ST Accel. Beams 8, 101001 (2005).

[2] V. Shiltsev et al., *Tevatron electron lenses: Design and operation*, Rev. ST Accel. Beams 11, 103501 (2008).

[3] J. Millman, H. Taub, Pulse, Digital and Switching Waveforms, McGraw-Hill, Inc., New York, 1965, pp. 69-72.

[4] Ibid., pp. 72-73.